\documentstyle[mprocl]{article}

\def\be{\begin{equation}}
\def\ee{\end{equation}}
\def\bea{\begin{eqnarray}}
\def\eea{\end{eqnarray}}
\begin{document}

\title{Can Spontaneous Supersymmetry Breaking in a Quantum Universe \\
Induce the Emergence of Classical Spacetimes ?}

\author{O. BERTOLAMI}

\address{Departamento de Fisica, IST \\
Av. Rovisco Pais, 1000 Lisboa, Portugal}

\author{P.V. MONIZ~ \footnote{Speaker}}

\address{{DAMTP, 
University of Cambridge\\ Silver Street, Cambridge, CB3 9EW, UK
\\{\small e-mail: {\sf 
prlvm10@amtp.cam.ac.uk; 
paul.vargasmoniz@ukonline.co.uk}}
\\{\small WWW-site: {\sf 
http://www.damtp.cam.ac.uk/user/prlvm10}}}}

\maketitle\abstracts{We introduce the set of constraints the wave
function of the Universe has to satisfy in order to describe an Universe 
undergoing through the process of spontaneous breaking of supersymmetry    
and discuss the way this may lead to the emergence of classical spacetime.}

{\em Can the  presence of supersymmetry introduce any significative changes 
in understanding the process of retrieval of classical behaviour in 
the framework of quantum cosmology ?}  In this contribution we shall address 
the issue of retrieving classical properties from supersymmetric 
quantum cosmologies, starting firstly with a brief review of  
the situation in standard quantum cosmology. 

In standard quantum cosmology, conserved probability 
currents can be obtained by requiring the wave function of the 
universe to be  of the form $\Psi_{WKB} \sim e^{iS}$, where $S$ is the 
classical action. 
As consequence, classical properties of specetime do emerge from $\Psi_{WKB}$. 
But what does one obtain from $e^{S_{E}}$, where $S_{E} = i S$ ?
In the case the wave function is an exponential 
rather than an oscillating function, 
$S_{E}$ corresponds to the action of  an 
Euclidean instead of a Lorentz  geometry. This is the situation when no
matter is present and the dominant saddle-point contribution to the 
path--integral is a real Euclidean 
solution of the field equations, a 
conclusion that holds for a variety of homogeneous 
minisuperspace models \cite{swh}.
However, it is important to notice that in this case
the wave functional $e^{S_{E}}$ is {\em not} peaked around a 
set of Euclidean solutions as it predicts no classical correlations between 
bosonic coordinates and momenta. In contrast, an oscillating wave function 
$e^{iS}$  is peaked around a set of classical Lorentzian trajectories
\cite{jjh}.

In supersymmetric quantum cosmology, on the other hand, 
most of the known solutions \cite{pmon} include {\it 
only} the exponential of the Euclidean action
$e^{\pm S_{E}}$, implying that they do {\em not} 
induce any classical Lorentzian geometry. 
This means that the supersymmetric minisuperspace models that 
have been currently studied 
still require additional elements in order to give origin to 
oscillating $e^{iS}$ solutions. 

In the remaining of this report we shall point out 
that the presence of a potential $V(\phi, \bar \phi)$ in the supergravity
action, where $\phi$ and $\bar \phi$ are chiral
superfields, may induce the transition 
from a supersymmetric quantum cosmological Euclidean 
phase into a classical Lorentzian inflationary expansion regime
where supersymmetry is broken. 
Naturally, such a potential leads to a complicate mixing between the 
fermionic sectors of the wave function as can be seen from
the constraint equations below. 
Furthermore, such a potential is related to a  
{\em  superpotential} $P[\phi, \bar \phi]$ and implies in the
possibility of {\sf  Spontaneous Supersymmetry Breaking} 
(SSSB), namely when, in supergravity, the v.e.v. 
of the K\"ahler derivative is non-vanishing
\cite{vnn}:
\begin{equation}
<D_{\phi} P> \equiv <\frac{\partial{P}}{\partial{\phi}} + 
P\frac{\partial{K}}{\partial{\phi}}> \not= 0~~,
\end{equation}
where the K\"ahler function is given by 
$K = \phi \bar \phi$ and we have set 
$M \equiv \frac{M_{Planck}}{\sqrt{8 \pi}}$ to 1
(see ref. \cite{OO} for a 
different presentation of SSSB in canonical supergravity).

As mentioned in the introduction the aim of our research is relating 
elements intrinsic to SSSB with the retrieval of 
classical properties from the wave function, $\Psi_{SUSY}$, 
corresponding to actions dominated by Lorentzian solutions \cite{bmonnew}.
We consider the simple case of a closed supersymmetric 
Friedmann-Robertson-Walker (FRW) universe with 
scalar fields and a superpotential of the form
\begin{equation}
 P = \lambda\phi\bar{\phi}~~,
\end{equation}
where in order to satisfy the phenomenology, $\lambda << M$.
The wave function of the universe takes the well 
known form \cite{pmon}
\begin{equation}
\Psi = A + B\psi^F \psi_F + C\psi^F \chi_F 
+ D \chi^F\chi_F + E \psi^F \psi_F \chi^G \chi_G~~, 
\label{eq:expansion}
\end{equation}
with $A, B, ...$ being the bosonic amplitudes 
corresponding to each fermionic sector and $F, G$ 
being spinor indices.
For our FRW model we obtain 
the following  set of equations:
\begin{eqnarray}
\frac{\partial{A}}{\partial{\phi}} &  + &
\sqrt{2}a^{3} \exp^{\frac{\phi\bar{\phi}}{2}}
\overline{D_{\phi} P}~D 
+ \frac{\sqrt{3}}{2}a^{3} e^{\frac{\phi\bar{\phi}}{2}}
\bar{P} C = 0, \label{eq:fa}\\ 
\frac{a}{\sqrt{3}} \frac{\partial{A}}{\partial{a}} 
+ 2\sqrt{3}a^{2}A & + &
2\sqrt{3}a^{3} e^{\frac{\phi\bar{\phi}}{2}} \bar{P}B 
+ \sqrt{2}a^{3} e^{\frac{\phi\bar{\phi}}{2}}
\overline{D_{\phi} P}~C = 0, \label{eq:fb}\\
\frac{\partial{B}}{\partial{\phi}} &  + &
\frac{1}{2} \bar \phi B - \frac{a}{4\sqrt{3}} 
\frac{\partial C}{\partial a} - \frac{\sqrt{3}}{2}a^2 C + \frac{C}{4\sqrt{3}}
\nonumber \\ & + &  
\sqrt{2}a^{3} \exp^{\frac{\phi\bar{\phi}}{2}}
\overline{D_{\phi} P}~E = 0 \label{eq:fc} \\ 
\frac{a}{\sqrt{3}} \frac{\partial{D}}{\partial{a}} 
& + &  2\sqrt{3}a^{2}D 
- \sqrt{3} D  + 
\frac{\partial{C}}{\partial{\phi}}   + 
\frac{1}{2} \bar \phi C \nonumber \\
& + & 2\sqrt{3}a^{3} e^{\frac{\phi\bar{\phi}}{2}} \bar{P}E = 0,
\label{eq:fd}
\\
\frac{\partial{E}}{\partial{\bar \phi}} &  + &
\sqrt{2}a^{3} \exp^{\frac{\phi\bar{\phi}}{2}}
{D_{\phi} P}~B 
+ \frac{\sqrt{3}}{2}a^{3} e^{\frac{\phi\bar{\phi}}{2}}
P ~C = 0, \label{eq:fe}\\ 
\frac{a}{\sqrt{3}} \frac{\partial{A}}{\partial{a}} 
& - &  2\sqrt{3}a^{2}A - 
\sqrt{3}a^{3} e^{\frac{\phi\bar{\phi}}{2}} P ~D 
+ \frac{1}{\sqrt{2}}a^{3} e^{\frac{\phi\bar{\phi}}{2}}
{D_{\phi} P}~C = 0, \label{eq:ff}\\
\frac{\partial{D}}{\partial{\bar \phi}} &  + &
\frac{1}{2}  \phi D +  \frac{a}{4\sqrt{3}} 
\frac{\partial C}{\partial a} - \frac{\sqrt{3}}{2}a^2 C -  \frac{C}{4\sqrt{3}}
\nonumber \\ & - &  
\sqrt{2}a^{3} \exp^{\frac{\phi\bar{\phi}}{2}} 
{D_{\phi} P}~A = 0 \label{eq:fg} \\ 
\frac{a}{\sqrt{3}} \frac{\partial{B}}{\partial{a}} 
& -  &   2\sqrt{3}a^{2}B 
- \sqrt{3} B   -  
\frac{\partial{C}}{\partial{\bar \phi}}   + 
\frac{1}{2} \phi C \nonumber \\
& -& \sqrt{3}a^{3} e^{\frac{\phi\bar{\phi}}{2}} \bar{P} ~A = 0.
\label{eq:fh}
\end{eqnarray}

The main features of our approach is the following. 
Introducing  a perturbative expansion in terms of $\lambda$
such as 
\begin{equation}
A = A_{0} + \lambda A_{1} + \lambda^{2} A_{2} + ...~~,
\end{equation}
and similarly for the other bosonic coefficients,
we get from (\ref{eq:fa}) and (\ref{eq:fb}), 
equations of the type
\begin{eqnarray}
\frac{\partial{A_{0}}}{\partial{\phi}}
& + & \lambda\left[ 
\frac{\partial{A_{1}}}{\partial{\phi}} + 
\sqrt{2}a^{3} e^{\frac{\phi\bar{\phi}}{2}} 
\overline{D_{\phi} P}D_{0} + 
\frac{\sqrt{3}}{2}a^{3} e^{\frac{\phi\bar{\phi}}{2}} 
\bar{P}~C_{0} \right] 
\nonumber \\ & + &  
\lambda^2 \left[\sqrt{2}a^{3} e^{\frac{\phi\bar{\phi}}{2}} 
\overline{D_{\phi} P}~D_{1} + 
\frac{\sqrt{3}}{2} ~a^{3} e^{\frac{\phi\bar{\phi}}{2}} 
\bar{P}C_{1}\right] = 0, \label{eq:ga} \\
\frac{a}{\sqrt{3}}\frac{\partial{A_{0}}}{\partial{a}} + 
2\sqrt{3}a^{2}A_{0} 
& + & 
\lambda \left[ 
\frac{a}{\sqrt{3}}\frac{\partial{A_{1}}}{\partial{a}} + 
2\sqrt{3}a^{2}A_{1} + 
2\sqrt{3}a^{3} e^{\frac{\phi\bar{\phi}}{2}}\bar{P}B_{0} 
+ \sqrt{2}a^{3} e^{\frac{\phi\bar{\phi}}{2}} 
\overline{D_{\phi}P}~C_{0}  \right] \nonumber \\ 
& + &  
\lambda^2 
\left[ 2\sqrt{3}a^{3} e^{\frac{\phi\bar{\phi}}{2}}
\bar{P}~B_{1} + 
\sqrt{2}a^{3} e^{\frac{\phi\bar{\phi}}{2}} 
\overline{D_{\phi}P}~C_{1}\right] = 0~~, \label{eq:gb}
\end{eqnarray}
up to $\lambda^2$-order. 
There are, of course, similar expressions 
for the remaining bosonic coefficients. 
The main point however, is that the bosonic coefficients 
$A_0, ..., E_0$ have already been determined \cite{pmon}. 
This is to say that, the equations they satisfy, which 
correspond to terms of order $\lambda^0$, must be equated to zero. 
Hence, neglecting terms in $\lambda^2$ allow us to 
solve the constraint equations in terms of 
$\lambda$. It will be then this perturbative solution of 
the constraint equations that enables us 
to address the issue of the retrieval of classical features
and relate it with the SSSB \cite{bmonnew}.

\section*{Acknowledgments}
PVM research work was supported by the JNICT/PRAXIS XXI Fellowship 
BPD/6095/95. The authors wish to thank 
M. Cavagli\`a, G. Esposito, C. Kiefer, S.W. Hawking, 
R. Graham, H. Luckock  and A. Vilenkin for useful conversations.

\end{document}